# Robotic observation pipeline for small bodies in the solar system based on open-source software and commercially available telescope hardware

Tobias Hoffmann[1,2]\*, Matti Gehlen[1], Thorsten Plaggenborg[3], Gerhard Drolshagen[1], Theresa Ott[1], Jutta Kunz[2], Toni Santana-Ros[4,5], Marcin Gedek[6], Rafał Reszelewski[6], Michał Żołnowski[6] and Björn Poppe[1]

[1]Department of Medical Physics and Acoustics, University of Oldenburg, Oldenburg, Germany, [2]Institute of Physics, University of Oldenburg, Oldenburg, Germany, [3]Faculty of Mathematics and Science, University of Oldenburg, Oldenburg, Germany, [4]Departamento de Física, Ingeniería de Sistemas y Teoría de la Señal, Universidad de Alicante, Alicante, Spain, [5]Institut de Ciències del Cosmos, Universitat de Barcelona, Barcelona, Spain, [6]6 Remote Observatories for Asteroids and Debris Searching (6ROADS), Cracow, Poland

The observation of small bodies in the Space Environment is an ongoing important task in astronomy. While nowadays new objects are mostly detected in larger sky surveys, several follow-up observations are usually needed for each object to improve the accuracy of orbit determination. In particular objects orbiting close to Earth, so called Near-Earth Objects (NEOs) are of special concern as a small but not negligible fraction of them can have a non-zero impact probability with Earth. Additionally, the observation of manmade space debris and tracking of satellites falls in the same class measurements. Telescopes for these follow-up observations are mainly in a aperture class between 1 m down to approximately 25 cm. These telescopes are often hosted by amateur observatories or dedicated companies like 6ROADS specialized on this type of observation. With upcoming new NEO search campaigns by very wide field of view telescopes, like the Vera C. Rubin Observatory, NASA's NEO surveyor space mission and ESA's Flyeye telescopes, the number of NEO discoveries will increase dramatically. This will require an increasing number of useful telescopes for follow-up observations at different geographical locations. While well-equipped amateur astronomers often host instruments which might be capable of creating useful measurements, both observation planning and scheduling, and also analysis are still a major challenge for many observers. In this work we present a fully robotic planning, scheduling and observation pipeline that extends the widely used open-source cross-platform software KStars/Ekos for Instrument Neutral Distributed Interface (INDI) devices. The method consists of algorithms which automatically select NEO candidates with priority according to ESA's Near-Earth Object Coordination Centre (NEOCC). It then analyses detectable objects (based on limiting magnitudes, geographical position, and time) with preliminary ephemeris from the Minor Planet Center (MPC). Optimal observing slots during the night are calculated and scheduled. Immediately before the measurement the accurate position of the minor body is recalculated and finally the images are taken. Besides the detailed description of all components, we will show a complete robotic hard- and software solution based on our methods.

# 1 Introduction

Follow-up observations and tracking of fast-moving small objects in the Space Environment are an important task in modern astronomy. Scientific knowledge can be gained in particular from a deeper understanding of the dynamics of small bodies in the Solar System and from the interaction with the gravitational influences of the Sun and the planets (Koschny et al., 2017). Furthermore, potential impacts of sufficiently large objects pose a danger for the Earth or space missions (Rumpf et al., 2016). Due to the immense increase of space missions in the last years (new-space-era) also follow-up observations of space-debris play a more and more important role in this field of astronomy.

Objects that may come close to Earth are called *Near-Earth Objects* (NEOs). In 2005, NASA was given a mission to find and track 90% of all Potentially Hazardous Asteroids (PHA) by the end of 2020. The significantly increased efforts resulted in an exponential increase in the number of objects discovered in the following years.[1] Ground-based search campaigns like the Vera C. Rubin Observatory or the Flyeye telescope (ESA) and space missions like the NEO surveyor space mission (NASA) are expected to increase the number of discovered objects in the next years significantly (Cibin et al., 2019; Mainzer et al., 2021). Data are collected internationally in a centralized manner by the *Minor Planet Center* (MPC) and made available for further evaluations.[2]

The detection and tracking of the observable, generally called "small objects," is therefore a key challenge for international risk assessment and planetary defense (Rumpf et al., 2016; Mainzer et al., 2021). To support the task of follow-up observations, observatories all around the world observe Minor Planets (Figure 1). Typically, telescopes up to 2 m in diameter are used here.

However, the number of available observatories is still too small compared to the necessary number of follow-up observations. Amateur astronomers often operate small observatories, whose instruments would in principle be able to perform meaningful measurements. Even though some sites are already submitting measurements to the MPC, a large part of this potential remains unused. Among other things, we see one of the major reasons in the relatively complicated object selection, where many parameters (ephemerides, brightness, location of the observatory, limiting magnitude, ...) must be considered, some of which change dynamically, in order to carry out a successful measurement.

However, many robotic telescope systems and networks for NEO follow-up observations also exist (e.g., Las Cumbres Observatory Global Telescope (LCOGT, (Shporer et al., 2010))) or are under construction (Lister et al., 2015; Dotto et al., 2021), but there are no commercially available solutions for amateur or small professional equipment (e.g., for university observatories) so far for that specific use, if then only in parts (Gupta et al., 2015; García-Lozano et al., 2016). In the recent years, powerful software and hardware enhancements have been developed, especially for the amateur sector, which make it possible to operate even small observatories worldwide robotically or at least remotely.

In this work we will describe a viable modular solution with commercially available hardware and software that makes it possible to perform follow-up observations of Minor Planets and thus contribute to their improved orbit prediction. The robotic system consists of a planning, scheduling and observation pipeline that is based on the open-source software KStars/Ekos with INDI devices.[3] It automatically obtains the objects and their position data from ESA and the MPC, constantly makes updates on the ephemeris, controls the observatory and automatically schedules different objects for observing nights.

# 2 Observational targets

## 2.1 Small Solar System bodies

The International Astronomical Union (IAU) defines small bodies of the Solar System as all objects apart from planets, dwarf planets and (natural) satellites orbiting the Sun (IAU General Assembly, 2006). Therefore, all Comets and Minor Planets (without dwarf planets) belong to this category. Within this group Asteroids and Comets are the major types of Small Solar System Bodies (SSSB). Due to their low mass, their physical and orbital behavior like the shape and the orbit stability can be quite different from planets.

---

1 https://cneos.jpl.nasa.gov/stats/totals.html
2 www.minorplanetcenter.net/iau/mpc.html
3 www.indilib.org/what-is-indi/discover-indi.html

The study of these objects is of great interest because gravitational and non-gravitational perturbations and collisions may lead to a change of orbit (Bottke et al., 2006). It turns out that the size distribution follows an inverse power law, so there are far more smaller objects than larger ones, which makes most of them quite difficult to detect because they are so faint (Peña et al., 2020). Since SSSBs are either remnants from the formation of the Solar System or fragments of a collision, it makes them scientifically relevant for the understanding of our Solar System (Hestroffer et al., 2019).

## 2.2 Near-Earth Objects

The movement of the SSSBs is highly dynamic. Up to date, we know more than 1 million Minor Planets, the majority of them orbiting between Mars and Jupiter in the Main Belt. However, orbital resonances and disturbances can bring some of these objects to inner orbits approaching Earth. SSSBs that have a perihelion of less than 1.3 AU are defined as Near-Earth objects (NEOs). The term Near-Earth Asteroid is also used because most NEOs are Asteroids and only a small fraction are Comets. So far, nearly 30.000 Near-Earth Asteroids have been discovered, with increasing numbers daily.[4] The minimum orbit intersection distance (MOID) is the minimum distance between the orbits of two objects. If such an NEO has a size of more than 140 m and a MOID of less than 0.05 AU to the Earth's orbit, it is classified as a Potentially Hazardous Object (PHO, (Huebner et al., 2009)). Currently, there are more than 2.000 of these PHO known.[5] Even if being smaller, some NEOs can be a threat for our planet. For example, the Chelyabinsk meteor event in 2013, which was an asteroid with an estimated effective diameter of about 18 m, led to many injured people.[6] According to current research, such an event is possible about every 50 years (Boslough et al., 2015).

To detect such objects early in advance, sky surveys for NEOs like the Panoramic Survey Telescope & Rapid Response System (Pan-STARRS) or the Catalina Sky Survey are regularly scanning the sky for new objects (Larson et al., 1998; Hodapp et al., 2004).

## 2.3 Minor Planet Center

The Minor Planet Center (MPC), which is under the guidance of the International Astronomical Union (IAU), is in charge of the worldwide organization of all data of SSSBs, especially NEOs and PHOs.[7] It collects astrometric and photometric data from the individual observatories, combines and processes them to calculate the orbits.

Every observatory that contributes positional data of Minor Planets to the MPC has an assigned individual code, consisting of three characters, in the form of a combination of letters and numbers. This MPC code is needed in advance to submit the data. It is assigned to an observatory when its initial submission with specified requirements is accepted.[8] The positional accuracy of the submitted objects is expected to be within two arcsec compared to the predicted ephemeris. There is a specific format for reporting the measurements, which is already implemented in some evaluation software (e.g., *Astrometrica*[9]).

## 2.4 NEO search campaigns

NEO candidates, that have recently been discovered, require confirmation by follow-up observations of other observatories (Micheli et al., 2015). They allow to confirm that the object is real and has the appropriate perihelion distance (Seaman et al., 2021). Further measurements will then be needed to improve the accuracy of the ephemeris, which is needed for further investigations on the object's physical properties. NEOs are usually only observable during their close approach to Earth, meaning that they can be lost if their ephemeris has a large uncertainty due to observations with low precision or only from a small orbital arc (Micheli et al., 2014). Additionally, perturbation by gravitational forces of other objects or non-gravitational forces like absorption and emission of radiation are increasing the uncertainty (Bottke et al., 2006; Perna et al., 2013). The targeted recovery of such lost object is so difficult, such that it rather will be found again by ordinary survey observations (Milani, 1999). This shows that sufficient follow-up observations over as long a period as possible are of importance to minimize the risk for lost objects and clarify the future impact possibilities (Micheli et al., 2014).

The MPC recommends observers to generally make two or three measurements per object per night.[10] It is sufficient if the observations are made over the period of a few hours. To remove ambiguity, this procedure should be repeated on another nearby night. It is explicitly not necessary to make more than three measurements for one object per night. For a potential new discovery, it is reasonable to make measurements on groups some hours apart on a single night. For follow-up observations of a new discovery, it is recommended to make measurements on pairs of nearby nights for every seven to 10 days repeatedly as long as the object is visible and unidentified. So, it is not necessary to observe a new object each night. If an object has already passed through several oppositions, measurements should be made on pairs on nearby nights around each opposition.

---

4 https://cneos.jpl.nasa.gov/stats/totals.html
5 https://cneos.jpl.nasa.gov/stats/totals.html
6 https://cneos.jpl.nasa.gov/news/fireball_130301.html

7 www.minorplanetcenter.net/iau/mpc.html
8 www.minorplanetcenter.net/iau/info/Astrometry.html
9 www.astrometrica.at
10 www.minorplanetcenter.net/iau/info/Astrometry.html

TABLE 1 Components and specifications of the telescopes "GHOST" and "ORT" of the University Observatory of Oldenburg (MPC Code: G01).

| Telescopes | *GHOST* | *ORT* |
| --- | --- | --- |
| *OTA* | 16-inch, 3250 mm (f/8) Ritchey-Chrétien Telescope | 6-inch, 420 mm (f/2.8) Corrected hyperbolic Astrograph |
| *Mount* | High-precision GoTo GEM with absolute encoders | GoTo GEM |
| *Focus* | Temperature compensated | Temperature compensated |
| *Filters* | 7 × 2-inch: L, R, G, B, SII, Ha, OIII | 8 × 1.25-inch: L, Photometric BVRI, SII, Ha, OIII |
| *Camera* | Cooled b/w CCD camera, 17.6 mm × 13.52 mm, 5.4 µm pixels | Cooled b/w CMOS camera, 13.2 mm × 8.8 mm, 2.4 µm pixels |
| *Field-of-View* | 0.31° × 0.24° | 1.80° × 1.20° |
| *Pixel scale* | 0.34"/pixel (1 × 1 Binning) | 1.18"/pixel (1 × 1 Binning) |
| *Add. Scope* | 50 mm, 205 mm (f/4.1) Refractor (for Alignment) | 50 mm, 190 mm (f/3.8) Refractor (for Guiding) |
| *Dome* | 3.0 m diameter, 0.5 m shutter controlled and powered wirelessly | 0.8 m diameter, two-part folding dome |

According to our analysis of the MPC database in 2021, for the initial discovery there are on average 21.9 ± 8.6 measurements needed to allow a clear and reliable classification. However, several (most ground-based) follow-up observations in much larger numbers in appropriate time intervals are usually necessary for sufficient orbit calculations over a longer period (Vereš et al., 2018).

## 2.5 Object selection

Due to the variety of observing strategies of NEOs, efficient methods are needed to make the best use of the observation time.[11] This can be done by prioritizing certain objects that benefit more from further observations than other (Micheli et al., 2015). A protocol was established for the ESA's NEOCC Priority List that selects NEOs for follow-up observations for which the improvement of orbital accuracy will be maximized with only minimal observing efforts (Boattini et al., 2007). Besides the MOID and the object's Sky Uncertainty (SU), an estimate of the difficulty to recover an object depending on the visual magnitude and further parameters are considered in this list. By combining these factors, the according urgencies are categorized into priority classes.

The Priority List can be accessed by an automated HTTP GET request.[12] The list contains additional information about the object's positions, physical characteristics, and orbital uncertainties. The individual entries can be used for further processing. For example, objects can be sorted out according to own requirements for urgency and visual magnitudes.

Besides the NEOCC Priority List there are also other ways to select suitable objects for observation nights. The NEA Observation Planning Aid (NEAObs) of the MPC makes a user-orientated approach and creates a list of suitable objects that fulfills the user's criteria (e.g., magnitude, motion, and uncertainty ranges) and are observable for the site.[13] The NEO Confirmation Page (NEOCP) contains the current NEO candidates that need confirmation for the discovery. Both can be accessed similar to the Priority List.

# 3 Instrumentation and implementation

## 3.1 Robotic Telescope instruments

The two Robotic Telescopes used in this work are located at the University Observatory of Oldenburg (08° 09' 55.0" E, 53° 09' 10.3" N, Elevation: 22 m above sea level, MPC Code: G01). The main telescope, called "**G**roßes **H**auptteleskop der **O**ldenburger **St**ernwarte" (GHOST), is a 16-inch f/8 Ritchey-Chrétien telescope used for fainter and small-sized astronomical objects. The smaller 6-inch f/2.8 corrected Astrograph, the "**O**ldenburg **R**obotic **T**elescope" (ORT), is used for fast wide-field imaging. Table 1 lists the detailed components and specifications of both telescopes. All hardware is commercially available.

The methods are developed and tested at the GHOST telescope in Oldenburg. The KAF8300 (onsemi[14], Phoenix, AZ, United States) chip of the camera is set in a 2 ×

---

11 https://neo.ssa.esa.int/priority-list
12 https://neo.ssa.esa.int/priority-list
13 www.minorplanetcenter.net/cgi-bin/neaobs.cgi
14 www.onsemi.com/

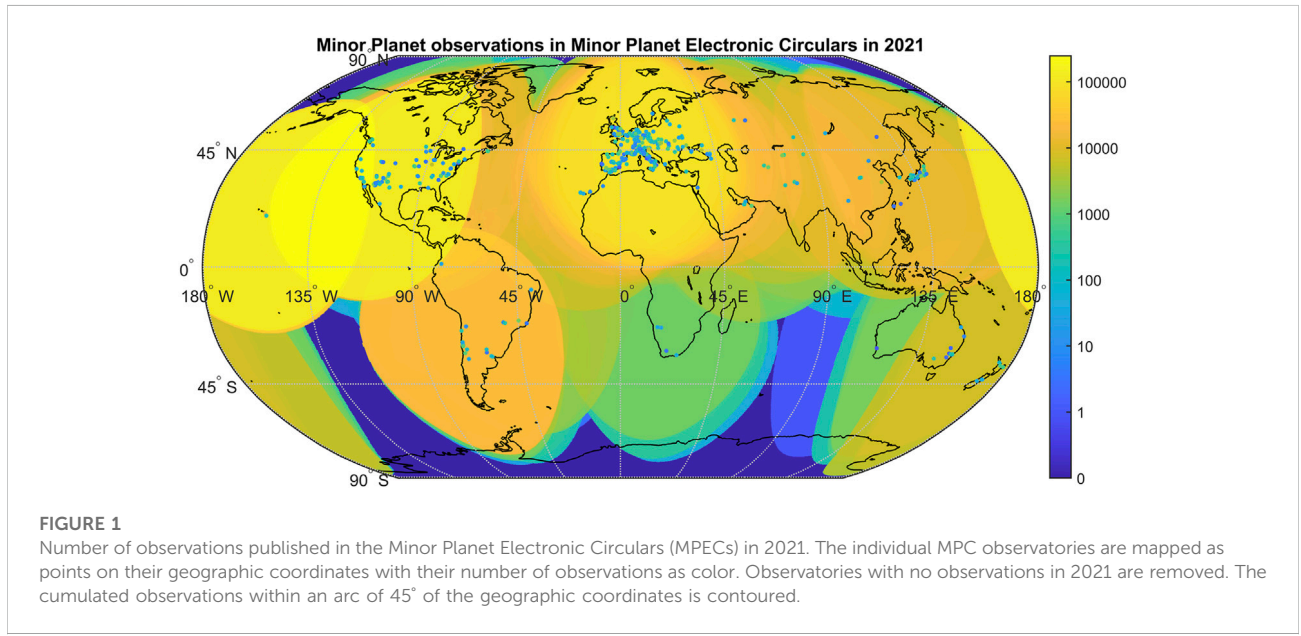

**FIGURE 1**
Number of observations published in the Minor Planet Electronic Circulars (MPECs) in 2021. The individual MPC observatories are mapped as points on their geographic coordinates with their number of observations as color. Observatories with no observations in 2021 are removed. The cumulated observations within an arc of 45° of the geographic coordinates is contoured.

2 Binning for NEO observation with an exposure time of $t_{exp}$ = 60 s. The bitrate is 16 bits with a full well of 25,000 $e^-$, the gain is g = 0.41 $e^-$/ADU (analog to digital unit) and the mean Quantum Efficiency is $QE$ = 0.54 (for wavelengths $\lambda$ = (550 ± 150) nm). The bias of the chip is measured to be $N_{bias}$ = 250 $e^-$. According to the chip's data sheet, the readout current is $N_{readout}$ = 7 $e^-$ and the dark current is 0.1 $e^-$/s at −10°C with a doubling temperature of 5.8°C. For other temperatures the dark current $N_{dark}$ at temperature $T$ can be calculated with:

$$N_{dark} = t_{exp} \cdot 0.1 \ \frac{1}{s} \cdot 2^{\frac{T+10°C}{5.8°C}} \qquad (1)$$

where a temperature of $T$ = −20°C is usually used. All the values $N$ refer to the corresponding number of electrons measured per pixel. The used clear glass filter has, according to its data sheet, a transmission of at least 98%. The main and secondary mirror of the telescope each have a dielectric high-reflectivity coating with at least 92% transmission. In total this results in a transmission of the optical tube of $\tau = 0.98 \cdot 0.92^2 \approx 0.83$. The secondary mirror itself has a diameter of 191 mm, thus resulting in a relative obstruction $a_{obstr}$ = 0.2213 of the light blocked for the primary mirror.

As an illustration of the number of expected follow-up observations, the presented methods are compared for different sites and observatories. For this purpose, we use the observatories of the 6ROADS network (Remote Observatories for Asteroids and Debris Searching)[15]. The network consists of six observatories with different sky qualities and telescope diameters/power classes. Additionally, the data for the ESA *Optical Ground Station* (OGS)[16] is added to the list since it is one of the most active telescopes in this field. Additional data on the telescopes are summarized in Table 2.

## 3.2 Software

The robotic telescope's software is based on *INDI Library* (Instrument-Neutral-Distributed-Interface) components. The system uses the INDI protocol to control the hardware, automate processes, collect data and exchange information among the devices and the software front-ends.[17] INDI consists of drivers to control the astronomical equipment of an observatory, a server as a central hub in between the drivers/devices and the clients/software, which can be accessed within the network, and the clients itself. One of the most generic GUI clients to control the devices is *KStars*. In Figure 2, the connection of the different INDI components from the client to the individual devices is schematically described. There are several more clients in INDI, like the *DBus Interface*, which can be used additionally to make automations for the observatory. It is fully scriptable and can control all devices with an interface. Together with an automated script, which also can connect by itself to the INDI server, but on a more fundamental way, it can be used for a Robotic control system.

---

15  https://6roads.com.pl/

16  https://sci.esa.int/web/sci-fmi/-/36520-optical-ground-station
17  www.indilib.org/what-is-indi/discover-indi.html

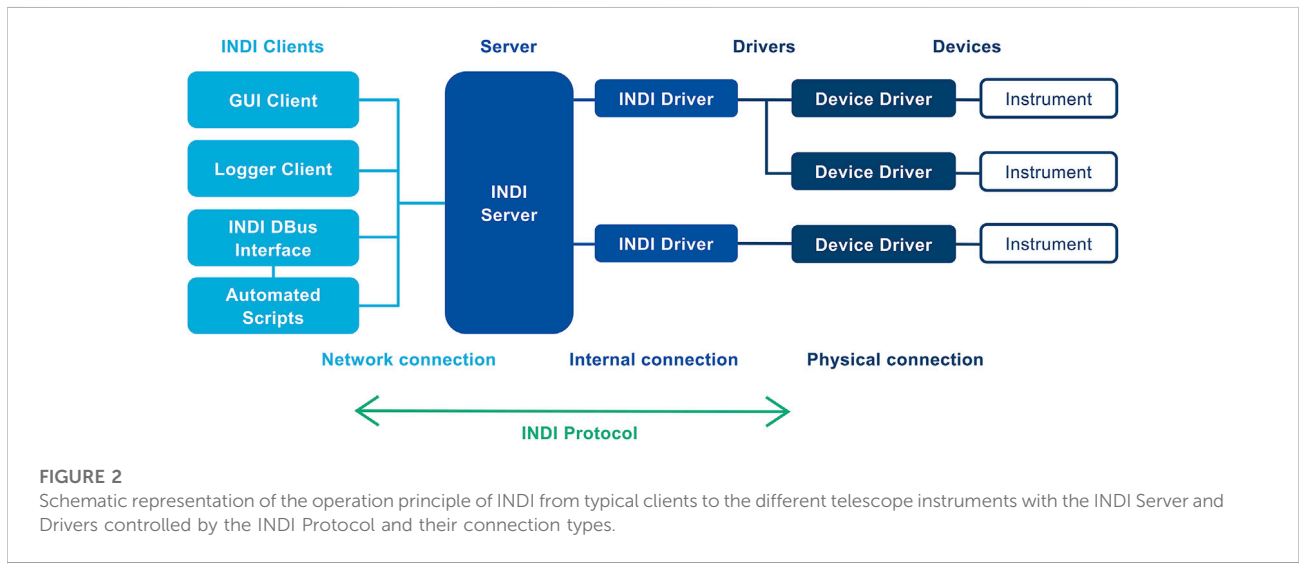

**FIGURE 2**
Schematic representation of the operation principle of INDI from typical clients to the different telescope instruments with the INDI Server and Drivers controlled by the INDI Protocol and their connection types.

*KStars* as one of the most common GUIs is an open-source cross-platform Astronomy Software.[18] In addition to the control over the INDI devices, it has capabilities for observation planning and graphical simulations of the night sky. The database contains up to 100 million stars, 13,000 deep-sky objects, all planets etc. Catalogs for Minor Planets or other objects can also be extended individually by adding external data from the MPC.

*Ekos* is the framework of KStars for data-acquisition and observatory control.[19] Due to its modular structure, it can be used for many automation processes. There are modules for the automatic capture-, focus-, mount slewing-, alignment- and guiding-process each. Additional accessories like weather monitoring and dome control can also be used in separate modules. All these modules can be automatically controlled *via* the Ekos Scheduler, which checks objects from a given list of targets for the current observing conditions and then makes a completely automatic and adjustable observation.

In addition to the optional startup and shutdown scripts, additional scripts can also be embedded before and after each observation of an object and each individual recording. This can be done by executable Python scripts. The INDI DBus Interface is implemented into Python.

After images are obtained by the robotic telescope system with Ekos, the positions and magnitudes of the object need to be analyzed. This astrometric and photometric data reduction can be done with the astrometry software *Astrometrica*.[20] The software can stack the images with a shift resulting from the expected movement of the object. For this, the orbital parameters from the MPC database are automatically obtained. This will improve the Signal-to-noise ratio and the distinction of the object from the stars as it prevents objects from becoming trails at long exposure times. The software then uses a Gaussian Point-Spread-Function (PSF) as fit function for data reduction in order to match reference stars (in our case from the Gaia DR2 catalog with the Gaia Broadband color band) in the picture (Raab, 2002). With that, the coefficients of the two-dimensional coordinate transformation, called plate constants, are calculated with a fourth-order polynomial fit. After extracting the astrometric and photometric measurements, the program allows to send a properly formatted MPC report.

## 3.3 Sky Brightness

For estimating the general brightness of the night sky a *Sky Quality Meter* (SQM) is used (Hänel et al., 2018). The detector measures the luminance in a field with a full width at half maximum (FWHM) of 20° near the zenith. The measurements are given in mag/arcsec$^2$, where this quantity corresponds approximately to the visual magnitude of the sky $m_{sky}$. A local deviation of the brightness can occur due to further influences (e.g. the Moon) and thus worsen the observation conditions. To minimize these influences, a minimum Moon distance of 30° is set.

With a field factor $F$, typically being a value between 1.4 and 2.4 according to the observer's visual capabilities and experience, Crumey (2014) derived the following expression for the limiting visual (naked eye) star magnitude $m_0$:

$$m_0 = 0.3834 \cdot m_{sky} - 1.4400 - 2.5 \log F \quad (2)$$

if $(20 < m_{sky} < 22)$ mag/arcsec$^2$. With $m_0$ a classification in the Bortle scale according to Bortle (2001) can be made.

---

18 https://edu.kde.org/kstars/
19 www.stellarmate.com/support/ekos/17-support/documentation/ekos.html
20 www.astrometrica.at

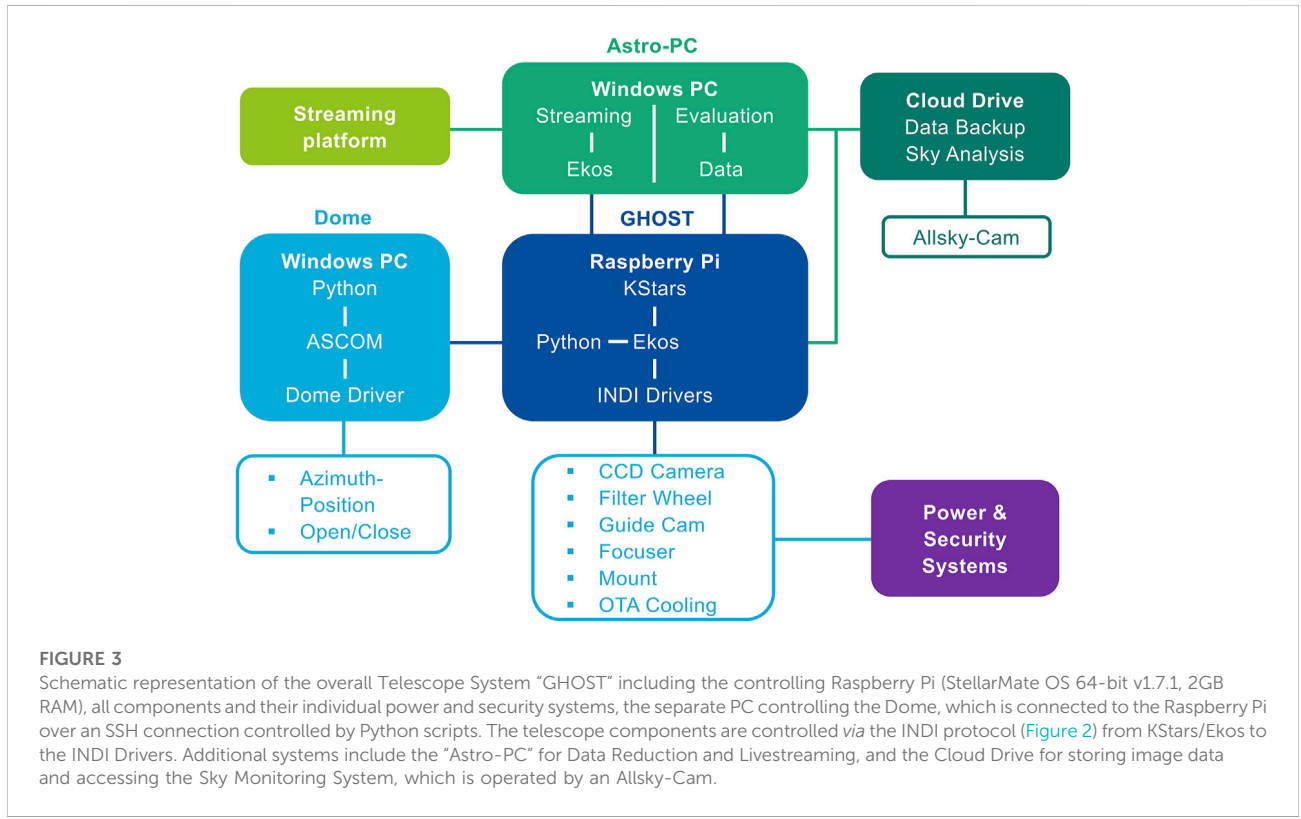

**FIGURE 3**
Schematic representation of the overall Telescope System "GHOST" including the controlling Raspberry Pi (StellarMate OS 64-bit v1.7.1, 2GB RAM), all components and their individual power and security systems, the separate PC controlling the Dome, which is connected to the Raspberry Pi over an SSH connection controlled by Python scripts. The telescope components are controlled *via* the INDI protocol (Figure 2) from KStars/Ekos to the INDI Drivers. Additional systems include the "Astro-PC" for Data Reduction and Livestreaming, and the Cloud Drive for storing image data and accessing the Sky Monitoring System, which is operated by an Allsky-Cam.

## 3.4 Estimation of imaging Limiting Magnitude

A crucial role for the automatic selection of observable objects is the imaging Limiting Magnitude (LM). It indicates up to which magnitude objects can still be observed and is mainly determined experimentally. However, it is useful to determine this value theoretically in order to estimate the influence of environmental and technical conditions. The LM depends on various telescope properties and settings. The following derivation is motivated on the calculations from Koschny and Igenbergs (2020). We start with the definition of the apparent magnitude different of two objects depending on the ratio of their flux densities Φ. When we consider one object as the Sun and the other as the observed object (e.g., NEO), the following is given:

$$m_{obj} = m_{Sun} - 2.5 \log\left(\frac{\Phi_{obj}}{\Phi_{Sun}}\right) \quad (3)$$

with the apparent magnitude $m_{obj}$ of the observed object, $m_{Sun}$ = −26.74 mag of the Sun[21] and the corresponding energy flux densities $\Phi_{obj}$ and $\Phi_{Sun}$ = 1361 W/m² (Mamajek et al., 2015). Since the spectrum of the Sun extends over a wide range, but the telescopes only measure the visible spectrum from 400 to 700 nm, the energy flux density is reduced to $\Phi_{Sun}$ = 535.5 W/m² (Meftah et al., 2018). For a given magnitude of the object its flux can therefore be calculated. With that, the total power of the signal $P_{cam}$ measured on the camera is calculated by multiplying the surface area of the telescope, depending on its diameter $d_{tele}$ of the primary mirror and its relative obstruction by the secondary mirror $a_{obstr}$, and a total transmission rate of the telescope $\tau$ with the energy flux density $\Phi_{obj}$:

$$P_{cam} = \Phi_{obj} \cdot \frac{\pi}{4} \cdot (d_{tele})^2 \cdot \tau \cdot (1 - a_{obstr}) \quad (4)$$

The integrated power of the signal over the exposure time $t_{exp}$ will give us the total energy of the photons captured by the camera. To convert this into the photoelectrons released in the camera sensor, we divide this by the energy per photon and multiply it with the quantum efficiency $QE$ of the camera. Assuming a mean wavelength of the photons $\bar{\lambda}$ and a constant power $P_{cam}$, we will get the number of electrons captured at the center of the signal $N_{signal}$:

$$N_{signal} = QE \cdot p \cdot \frac{P_{cam} \cdot t_{exp}}{hc/\bar{\lambda}} \quad (5)$$

with $p$ the percentage of the signal in the center pixel, Planck's constant $h$ and the speed of light $c$. Besides the signal to be measured, other disturbances occur. In addition, the readout noise $N_{readout}$, the thermal noise $N_{dark}$ and the camera's offset value $N_{bias}$ need to be considered. The main disturbance of the signal is the light from the empty sky $N_{sky}$ measured at the camera. This value depends significantly on the location (suburban, rural sky, *etc.*). To quantify this lightness, we can use the already known visual magnitude $m_{sky}$. Eq. 3 is analogously considered with magnitudes $m_{sky}$ and flux densities $\Phi_{sky}$ of the sky. Since the magnitude of the sky is given per arcsec$^2$, we need to multiply the resulting flux with the square area of one pixel with size $s_{px}$. Rearranged to the flux from the sky this results in:

$$\Phi_{sky} = s_{px}^2 \cdot \Phi_{Sun} \cdot 10^{\frac{m_{Sun}-m_{sky}}{2.5}} \tag{6}$$

---

21 https://nssdc.gsfc.nasa.gov/planetary/factsheet/sunfact.html

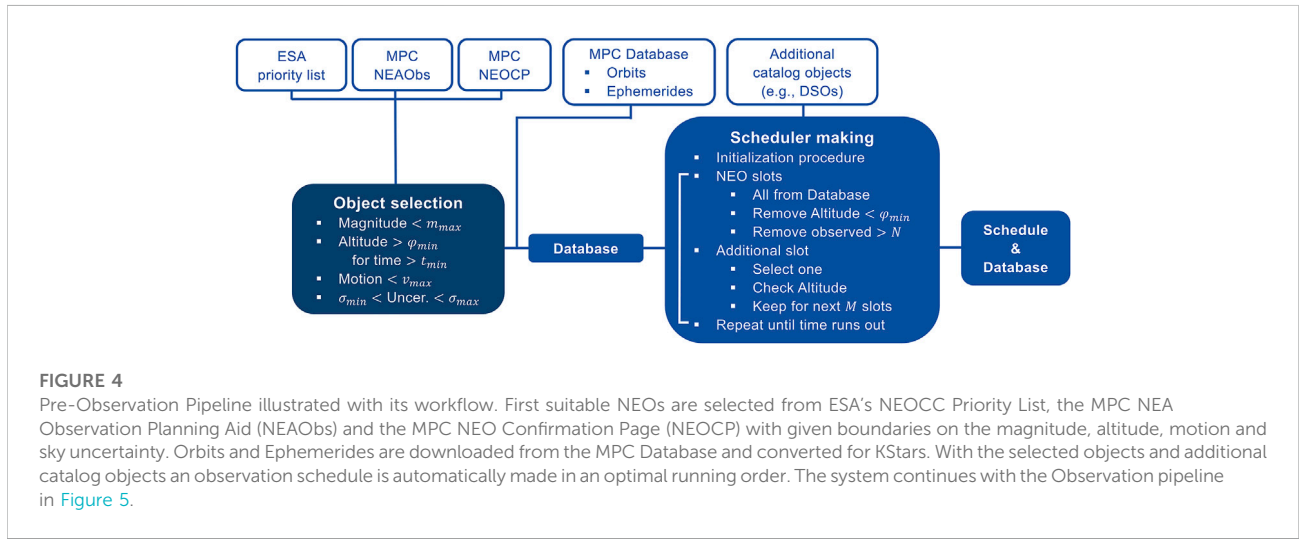

**FIGURE 4**
Pre-Observation Pipeline illustrated with its workflow. First suitable NEOs are selected from ESA's NEOCC Priority List, the MPC NEA Observation Planning Aid (NEAObs) and the MPC NEO Confirmation Page (NEOCP) with given boundaries on the magnitude, altitude, motion and sky uncertainty. Orbits and Ephemerides are downloaded from the MPC Database and converted for KStars. With the selected objects and additional catalog objects an observation schedule is automatically made in an optimal running order. The system continues with the Observation pipeline in Figure 5.

The Eqs 4, 5 can also be used analogously, where $p$ can be neglected, because the sky signal is equally distributed over the sensor. These results in the following:

$$P_{cam,sky} = \Phi_{sky} \cdot \frac{\pi}{4} \cdot (d_{tele})^2 \cdot \tau \cdot (1 - a_{obstr}) \tag{7}$$

$$N_{sky} = QE \cdot \frac{P_{cam,sky} \cdot t_{exp}}{hc/\bar{\lambda}} \tag{8}$$

Altogether, the signal-to-noise ratio $SNR$ as in Merline and Howell (1995) can be calculated:

$$SNR = \frac{N_{signal}}{\sqrt{N_{signal} + n_{px}\left(1 + \frac{n_{px}}{n_b}\right)\left(N_{bias} + N_{dark} + N_{sky} + N_{readout}^2 + g^2\sigma_f^2\right)}} \tag{9}$$

with $n_{px}$ pixels considered for measurement of the signal and $n_b$ pixels for the measurement of the background signal. The additional term $g^2\sigma_f^2$, with the gain $g$ and an estimate of the $1\sigma$-error by the A/D converter of the camera $\sigma_f \approx 0.289$, indicate the error from the A/D conversion (Merline and Howell, 1995). In order to get the LM of a telescope, we need to make the calculation in Eq. 9 backwards. We presume a certain threshold for $SNR$ and get the required object's signal value:

$$N_{signal} = \frac{SNR}{2}\left(SNR + \sqrt{SNR^2 + 4n_{px}\left(1 + \frac{n_{px}}{n_b}\right)\left(N_{bias} + N_{dark} + N_{sky} + N_{readout}^2 + g^2\sigma_f^2\right)}\right) \tag{10}$$

Using Eqs 3–5 we get for the magnitude $m_{obj}$ of the object for a given $SNR$:

$$m_{obj} = m_{Sun} - 2.5\left[\log\left(\frac{2hc}{\pi\Phi_{Sun}}\right) - \log\left(t_{exp}QE\tau\bar{\lambda}p(1-a_{obstr})(d_{tele})^2\right) + \log(SNR)\right.$$
$$\left. + \log\left(SNR + \sqrt{SNR^2 + 4n_{px}\left(1 + \frac{n_{px}}{n_b}\right)\left(N_{bias} + N_{dark} + N_{sky} + N_{readout}^2 + g^2\sigma_f^2\right)}\right)\right] \tag{11}$$

Finally for the limiting magnitude of the telescope, we need to correct the object's magnitude from Eq. 11 with the atmospheric extinction causing a dimming of the light by the terrestrial atmosphere. With the air mass $\chi$ and the extinction coefficient $\kappa$ this will lead to the following expression:

$$m_{lim} = m_{obj} - \kappa \cdot \chi \tag{12}$$

where, for an object in the zenith, $\chi = 1$ and $\kappa = 0.245$ for the V-band (Jurado Vargas et al., 2002). In addition, astronomical seeing, i.e. the broadening and blurring of point sources due to air turbulence in the atmosphere, is an important quantity affecting the LM. It attenuates the intensity of the light and thus increases the extinction coefficient $\kappa$ due to increased scattering of light. On the other hand, the widening of the recorded light spot from the object causes a decrease in the percentage $p$ of the signal in the center pixel. This gives us an expression for the LM of a telescope depending on its specifications and the requested $SNR$. Typically, a value of $SNR > 5$ is needed for the detection of an object.

## 3.5 Robotic Telescope system

Building on the existing INDI framework with KStars/Ekos as the interface, it was possible to create a robotic observation pipeline. All the observatory's equipment is connected to a small single-board computer (Raspberry Pi 4B) *via* USB (Focus, Filter Wheel, Cameras) and Network (Mount, Dome). The Dome's shutter opening and closing is controlled directly *via* a Bluetooth connection. Figure 3 shows a general illustration of the observatory's components and accessories connected to the controlling computer of GHOST. An analogous Setup has been used for the ORT telescope.

With the hardware set up, an appropriate software is necessary for the use of the Robotic Telescope. The difficulty is, that most available software is not developed for the measurement of NEOs due to their relative movement in the equatorial coordinate system. Many programs like KStars support the observation of planets and Minor Planets, but only well-known numbered objects. Since orbits are continuously updated, follow-up observations of unnumbered, new or even unconfirmed objects require frequent connection to the latest data. We solved this problem by using the customization options of Ekos in the form of embeddable scripts to ensure a continuous update of the coordinates of the objects before each observation.

## 3.6 Pre-observation pipeline

Figure 4 shows the Pre-Observation Pipeline developed for the Robotic Telescope. It starts with the selection of suitable objects. For that, ESA's NEOCC Priority List and MPC's NEOCP is accessed with a Python script at the beginning of the observation night. Also, a NEAObs list from the MPC will be produced for specified boundaries for the magnitude (brighter than $m_{max}$), the motion (below $v_{max}$) and the sky uncertainty (in-between $\sigma_{min}$ and $\sigma_{max}$). The Priority List and the NEOCP will be filtered by the script with the same boundaries as well. Additionally, the time for which the object's altitude is high enough (larger than $\varphi_{min}$) to be observable is calculated. Objects that fall under a given threshold $t_{min}$ for the time are sorted out. What remains is a list of suitable objects for observation.

For the determination of preliminary positions, the current orbital parameters and ephemerides of all objects are downloaded from the MPC Database and the MPC Ephemeris Service, respectively. After converting the parameters into the format of KStars, the objects are included into its database by adding them in the Asteroids file of KStars. Besides, a separate database with the ephemerides is created.

In the next step, an automated Observation Schedule is made for the Ekos Scheduler. For this purpose, an XML file containing all information for the Ekos Scheduler is created, i.e., all individual observation slots must be specified here in the appropriate order. In the beginning there is an initialization procedure used for testing all instruments and then doing an accurate auto-focusing after the camera sensor and the main mirror are cooled down (regular automatic refocusing can be set in the schedule and is also recommended). After that, the actual observation slots are assigned. All objects currently having an altitude below the threshold $\varphi_{min}$ are removed. This procedure is repeated for several runs with updated altitudes until the nighttime ends. Objects that have already been selected more than $N$ times are also removed. With that, a complete Observation Schedule for NEOs is made and the Pre-Observation procedure is finished.

## 3.7 Observation pipeline

For the actual Observation Pipeline, which is shown graphically in Figure 5, the schedule and the database are loaded into KStars/Ekos. After manually checking the schedule for correctness, the Ekos Scheduler can be launched. It begins with a Startup Script enabling the power supply of the instruments and checking the weather conditions. If the weather is safe, the dome's shutter will open. After a successful connection and security check of all components, the Startup Script is finished, and the Scheduler will continue with its built-in features (connecting the INDI Server to the devices, unparking dome and mount, *etc.*). Then, the actual Schedule starts with an initialization object (ideally some field near the zenith) to prepare the system (e.g., focusing). After that, Ekos performs each observation job with its selectable modules and settings (e.g., checking starting conditions, slewing mount, slaving the dome, focusing, aligning and guiding). The various modules communicate with the devices *via* the INDI protocol (Figure 2).

One major adaptation we made for the NEO observation is embedded in the optional pre-capture script. In this Python script the current best possible orbital parameters of the NEO are downloaded from the MPC Database and with that the current ephemerides are calculated immediately before each measurement. If there is some correction needed, the mount will adjust its position accordingly. This is done independently of Ekos with the INDI DBus Interface. With that, Ekos slews the mount to the approximate position and the DBus Interface will correct it with the latest information from the MPC. It turns out that such correction is useful not only for fast moving objects, but also slower objects due to the large time spans in an observation night. One of the main problems with the Ekos Scheduler is that it only uses the coordinates when the Schedule is created and does not adjust them according to the ephemeris. However, the procedure shown compensates for this limitation.

After all measurements are finished, the Ekos will park the mount and the dome and disconnect all devices. Then, the observation night ends with a Shutdown Script, which closes the dome and turns off all instruments and their power supply. The measured data are sent to a separate computer and can be evaluated. For educational and logging purposes the entire process combined with images of surveillance cameras is broadcasted over a livestream and is stored together with logs for error analyses.

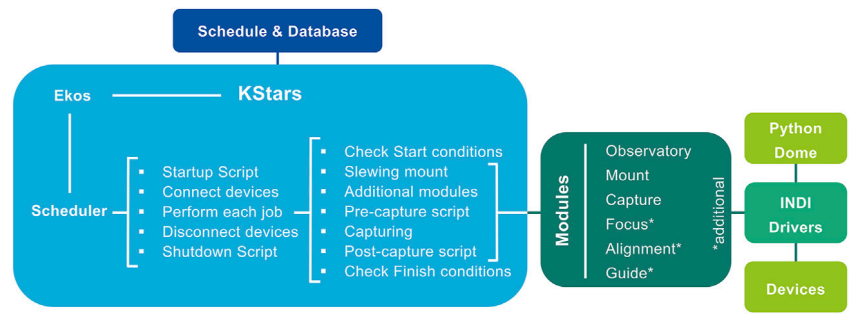

**FIGURE 5**
Observation Pipeline illustrated with its workflow. It stars with the Schedule and Database created by the Pre-Observation Pipeline in Figure 4 and starts the observation in the Ekos Scheduler. All operations from switching on the components through the capturing process to the end of the observation are executed automatically by the Scheduler functions. Here, customized scripts were embedded in the designated places at Startup, Shutdown and before and after each Capture respectively. When executing the jobs, the scheduler activates the corresponding Ekos modules, which in turn controls the devices *via* the INDI protocol.

# 4 Observations and reductions

## 4.1 MPC submissions

The described robotic system is regularly used for the observation of Minor Planets since Summer 2021. For the initial submission to the MPC in early July 2021, we made the measurements as listed in Supplementary Table S1A with in total 61 measurements of eight different Minor Planets, of which one was a NEO. Of those, 57 measurements were accepted by the MPC, which was enough for the assignment of the MPC code G01 for the observatory.

After that, 20 further reports have been made until mid-March 2022 as listed in Supplementary Tables S1B–U. In total 613 measurements were submitted to the MPC during this period with 570 measurements accepted. Of these, 384 submissions were from NEOs with 341 accepted (measurements may be rejected due to inconsistencies compared to measurements from other observatories).The rejected observations were re-measured, and about half of them were then accepted. The other half had erroneous measurements, so that objects other than the correspondingly indicated NEOs were measured by mistake. These were made on a total of 42 observation nights (27 with NEOs), some of which possessed only short observable phases. So, on average 14.6 measurements of Minor Planets in general and 14.2 of NEOs were made per night. It needs to be considered, that a single object was captured about two to four times per night.

The median of the measured magnitude of each object over the period of the report is calculated. The faintest magnitude is 19.4 mag for the objects 2022 CR3 and 2022 EB3 (Supplementary Tables 1T,U) using a 4 × 60 s stacked image, but actually the object 2022 EB3 were brighter with 19.0 mag according to the MPC. On the other hand, objects like 2022 DS4 with a measured magnitude of 18.9 mag were fainter with 19.5 mag, which is the LM for moving objects with these settings so far. For longer exposure times and slower moving objects, a higher LM can be achieved. On average, a value for the visual magnitude of (17.68 ± 0.93) mag was measured. Among the NEOs, the average value is (18.13 ± 0.57) mag.

In March 2022, the system also successfully discovered the new asteroid "2022 EX" with the preliminary data from other telescopes around the world in the NEOCP.[22] It was possible to make confirmatory and accurate measurements of the object within a short time after the initial discovery.

## 4.2 Sky Magnitude

We measured the Sky Brightness for the observation nights with an SQM-LU (Miguel et al., 2017).[23] The SQM is part of "Was het donker?" network of the university of Groningen.[24] The mean Sky Brightness at our site is (19.52 ± 0.48) mag/arcsec$^2$ with a faintest value of 20.40 mag/arcsec$^2$ measured.

Using Eq. 2 with a field factor $F = (1.70 ± 0.30)$ we can calculate the limiting visual star magnitude $m_0 = (5.48 ± 0.45)$ mag with the measured mean Sky Brightness. This results in a Bortle scale class of 5–6.

---

22 www.minorplanetcenter.net/mpec/K22/K22E59.html
23 www.unihedron.com/projects/darksky/cd/SQM-LU/SQM-LU_Users_manual.pdf
24 www.washetdonker.nl

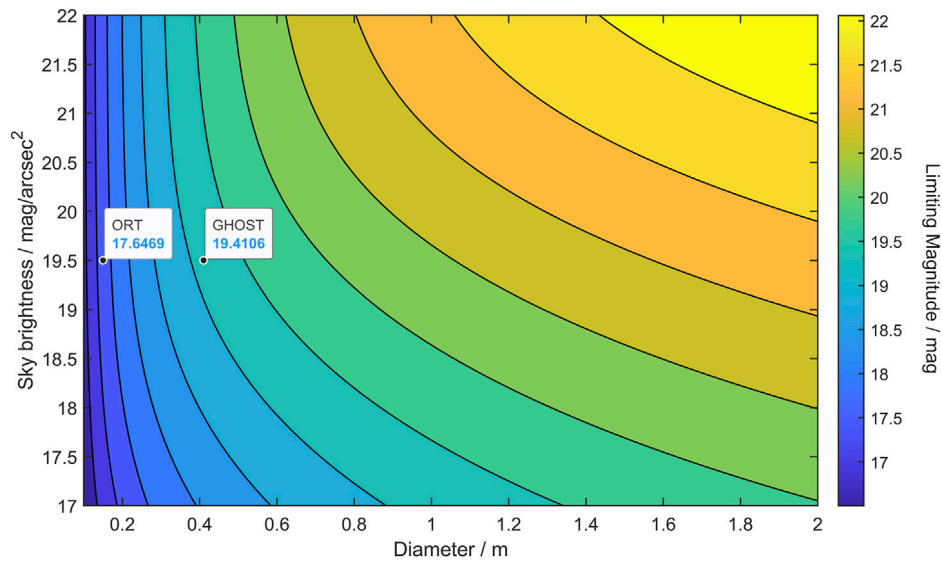

**FIGURE 6**
Estimation of the Limiting Magnitude for different telescope diameters and Sky Magnitudes calculated by Eq. 12 for an exposure time of 60 s with parameters based on the GHOST telescope (G01). Calculation results for the ORT and GHOST telescopes are marked in the graph.

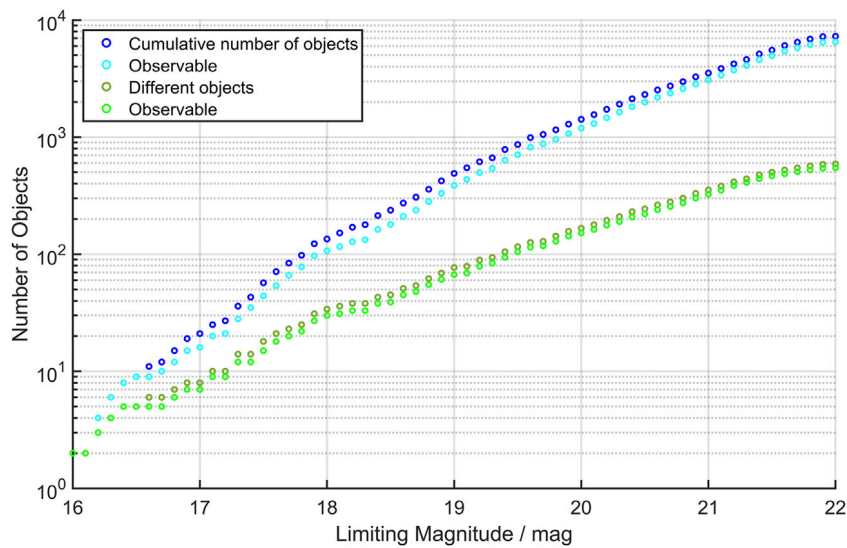

**FIGURE 7**
Number of objects in the NEOCC Priority List with a magnitude brighter than a particular Limiting Magnitude in the period from 01.12.2021 until 31.01.2022. The cumulative number of all entries (with multiple entries of the same object in different nights) and the observable of them (blue), and the different objects (counted only once for multiple entries on different nights) of these (green) are shown.

## 4.3 Observational Limiting Magnitude

In order to show, how many possible observations can be reached per night with our proposed methods for different telescope sizes and locations, we use Eqs 11, 12 to calculate an estimation for the LM. For this we assume a minimal $SNR$ of 5 for an object to be detectable and take as a basis the parameters of the GHOST telescope and constants from section 3.1 and section 3.4.

For the determination of the $SNR$, we consider only the center pixel of the signal ($n_{px} = 1$) with a background of $n_b =$

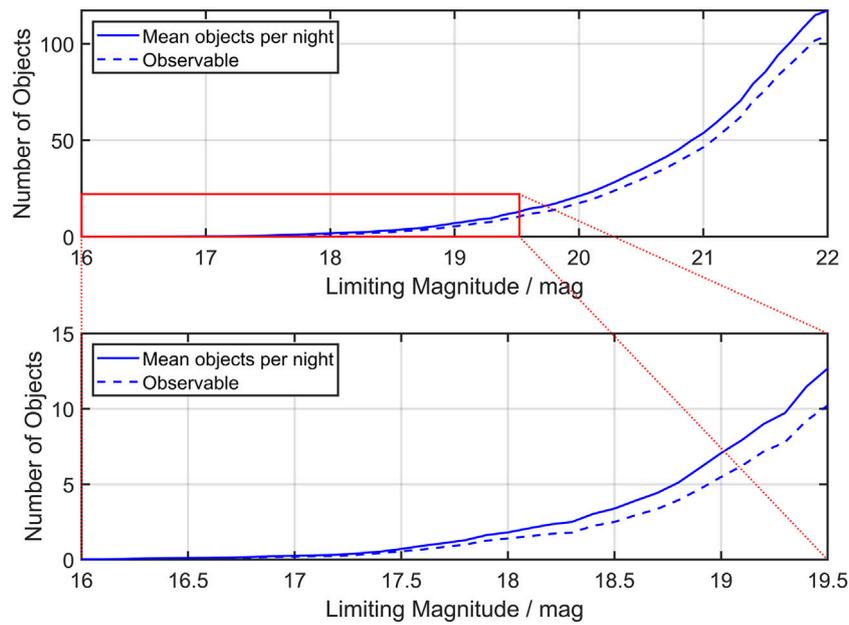

**FIGURE 8**
Results for the cumulative number of entries from the NEOCC Priority List from Figure 7 averaged per night with a magnitude brighter than a particular Limiting Magnitude and the observable of them (dashed). The range for the Limiting Magnitude from 16 to 22 mag is shown above, the extract of this up to 19.5 mag is shown below.

**TABLE 2** Telescopes in Oldenburg (GHOST, ORT), from the 6ROADS Network and ESA's Optical Ground Station (OGS) with their location, MPC Code and the results for the calculated Limiting Magnitude for the given diameter and Sky Magnitudes from Figure 6, the mean observable NEOCC Priority List object from Figure 8 and the mean observation time needed for those objects.

| Telescope | Location | MPC code | Diameter/m | Sky Mag./mag/arcsec² | Theo. LM/mag | Priority list objects | Obs. Time needed/h |
|---|---|---|---|---|---|---|---|
| ORT | Oldenburg, Germany | G01 | 0.15 | 19.5* | 17.6 | 0.7 | 0.18 |
| GHOST | | | 0.41 | | 19.4 | 9.2 | 2.30 |
| Solaris Obs | Cracow, Poland | B63 | 0.30 | 20.0 | 19.0 | 5.5 | 1.38 |
| Polonia Obs | San Pedro de Atacama, Chile | W98 | 0.25 | 22.0 | 18.8 | 4.0 | 1.00 |
| 6ROADS Obs. 1 | Wojnowko, Poland | K98 | 0.40 | 21.0 | 19.7 | 12.7 | 3.18 |
| Rantiga Obs | Tincana, Italy | D03 | 0.40 | 21.0 | 19.7 | 12.7 | 3.18 |
| Springbok Obs | Tivoli, Namibia | L80 | 0.36 | 22.0 | 19.6 | 11.9 | 2.98 |
| 6ROADS Obs. 2 | Nerpio, Spain | Z33 | 0.40 | 22.0 | 19.8 | 14.0 | 3.50 |
| ESA OGS | Tenerife, Spain | J04 | 1.00 | 21.5 | 21.4 | 70.2 | 17.55 |

The Sky Magnitude in Oldenburg (*) is measured with an SQM, all other values are obtained from Falchi et al. (2016).

100 pixels (10 × 10 box). Since the signal of the object observed is not focused on this one pixel due to external influences (e.g., seeing, deviations in focusing and collimation), we calculate the portion p of the signal in the center pixel with the mean Full Width at Half Maximum $\sigma_{FWHM}$ of the signal (assuming Gaussian intensity distribution). In our case a value of $\sigma_{FWHM} = (2.58 \pm 0.64)"$ was determined in Astrometrica. This includes the average seeing during the measurements. For a gaussian distributed signal $\sigma_{FWHM}$ is equal to $2\sqrt{2\log 2}$ times the standard deviation $\sigma$. By using the properties of the normal distribution, we can calculate $p$:

$$p = \text{erf}\left(\frac{2 s_{px} \sqrt{\log 2}}{\sigma_{FWHM}}\right)^2 = 0.145 \qquad (13)$$

with the Gauss error function erf and the pixel scale $s_{px} = 0.68"$/pixel (in binning 2 × 2). Now, for different values of the Sky Magnitude $m_{sky}$ we get the number of photons $N_{sky}$ captured at the camera with the Eqs 6–8. With the telescope's parameters but using different values for its diameter $d_{tele}$, Eq. 11 with the correction in Eq. 12 lets us estimate the LM $m_{lim}$ of different telescopes. The values of $m_{lim}$ for diameters $d_{tele}$ in range from 0.1 to 2 m and Sky Magnitudes $m_{sky}$ in range from 17.0–22.0 mag/arcsec² are shown in Figure 6. Using the diameter of GHOST ($d_{tele}$ = 0.41 m) and our measured mean Sky Brightness ($m_{sky}$ = 19.5 mag/arcsec²) we will get a theoretical LM of $m_{lim,GHOST}$ = 19.41 mag for a single 60 s exposure. With the values of ORT ($d_{tele}$ = 0.15 m, $m_{sky}$ = 19.5 mag/arcsec²) this will result in $m_{lim,ORT}$ = 17.65 mag. For longer exposures ($\alpha$ multiples of 60 s), Eq. 11 shows that the LM increases by a value of 2.5 log $\alpha$. With a typical 4 × 60 s stacked exposure this results in an increase of about 1.5 mag.

## 4.4 Estimating observational limits

The results for the LM can be used to estimate how many measurements of NEOs can be made per night. Due to the large number of NEOs detected, several of them can usually be observed in one night with a LM of about 17 mag.[25] Therefore, limited only by the total observation time, a large number of measurements can be made for LM fainter than that, but here we focus on observations of objects from the NEOCC Priority List. Such measurements can lead to significant improvement in orbital prediction and are therefore considered a priority (section 2.5).

We analyzed the Priority List for every night in December 2021 and January 2022 for the magnitudes of the objects. First, all entries of the different nights were cumulated and sorted by visual magnitude. Each entry is also checked for observability in terms of its altitude at the geographic coordinates of Oldenburg. Any object that was above 20° altitude for at least 2 h on the night indicated, is considered as observable.

Since objects may appear multiple times on the lists of different night, multiple entries are sorted out separately. Of all occurrences, the lowest value of the magnitude per object is used for the priority list. The cumulative number of all entries and the number of different objects that have a magnitude smaller than a limiting value are plotted in Figure 7. Results from the Priority List over a period of 62 days are used for estimates for individual nights. The resulting mean observable objects per night are plotted in Figure 8.

We can now use these results in connection with the estimations for the LM of different telescope sizes from Figure 6 and compute the mean observable Priority List objects for a given telescope diameter $d_{tele}$ and Sky Magnitude $m_{sky}$. The results for several observatories (our telescopes, 6ROADS Network and the ESA Optical Ground Station) are listed in Table 2. Assuming an measurement takes 5 min and three measurements per night are required for submission to the MPC, the total observation time required is also given.

## 5 Discussion

### 5.1 Robotic observation system

Due to the given customization possibilities in KStars, Ekos and INDI we were able to implement a robotic observation pipeline for Minor Planets and NEOs. Since the pipeline is an extension of the existing INDI framework the generalization of this approach to other observatories should easily be achievable. The pipeline itself is structured straightforward from the input of external data, the object selection to the observation procedure with modular and customizable options (Figures 3–5).

Many observations with our robotic system and the proposed methods were successful so far with increasing number (section 4.1 and Supplementary Table S1). The system delivers reasonable and reliable results and reduces the personnel effort immensely for the observation and planning.

But the pipeline can also be useful for other systems that are not robotic so far or are not specialized for NEOs. On one hand it is useful for larger telescopes, for which many objects requiring follow-up observations are available and thus need better efficiency for the observation time. On the other hand, there are amateur telescopes for which less objects are available, but more effort per capture is needed due to inaccuracies and the difficulty of the handling of the instruments. Therefore, such telescopes could also benefit from the simplicity of use of KStars and Ekos (e.g., its auto-guiding and aligning modules).

There are already successful Robotic Telescopes for NEO follow-up observations that automatically obtain data from the MPC Confirmation Page for the scheduling and update the ephemeris constantly (Holvorcem et al., 2003) or have developed highly automated pipelines (Dotto et al., 2021) and networks (Lister et al., 2015). There are also robotic solutions using commercial software (García-Lozano et al., 2016) and also with INDI, KStars and Ekos developed for small telescopes (Gupta et al., 2015), but there is currently no system that combines the usability and expandability of INDI with adaptions needed for NEO observations and the connectivity to the MPC and the NEOCC Priority List to realize an open-source Robotic Observation Pipeline that is applicable for commercially available instruments.

---

25 www.minorplanetcenter.net/whatsup/index

## 5.2 Limiting Magnitude and Priority List

We can analyze the estimations for the LM of the different telescope sizes and Sky Magnitudes with the results in Figure 6. For small telescopes up to 0.25 m it can be shown that the LM is relatively independent of the Sky Magnitude, which means that is negligible to some extent under which sky conditions the telescope is used. On the other hand, for larger telescopes above 1 m the LM is highly dependent of the sky conditions. This leads to the fact that smaller (and less expensive) telescopes with a good sky condition can make equal or better results as a large telescope with worse sky conditions. So it should be evaluated which telescope is optimal for a given location.

However, also smaller telescopes can obtain similar results if they spend more observation time per object. But since NEOs move relative to the reference stars during measurement, a longer exposure time is not always beneficial. Only a longer line trace of the object on the image is produced. Therefore, it is helpful to stack single measurements with shorter exposure times shifted by the movement of the object. With that, the LM for NEO observations can be increased. Such a stacking tool is implemented in the astrometry software *Astrometrica*.[26]

Figure 7 shows an exponential correlation of the number of objects in the NEOCC Priority List and the LM. This can be expected considering the power law for the absolute magnitude of these objects (Peña et al., 2020) and their distances. Therefore, the large number of faint objects require much observation time, but less telescopes are available for that.

While there is a small difference between the total number of objects and the observable ones, a big difference is noticeable between the cumulative number of entries and the number of different objects. The difference gets larger for higher LMs with a difference of more than one order of magnitude for objects fainter than 20 mag. This means that there are many objects, especially faint ones that remain on the list for many days (on average more than 10 days for objects fainter than 20 mag). This fits with the result that the many faint objects can hardly be covered with the few sufficient telescopes, which is a key result from a study by Seaman et al. (2021). However, a meaningful contribution can already be made with enhanced amateur telescopes with a LM above 18 mag. With these, measurements of several NEOs of the priority list can be performed in one night (Figure 8). Stacking can further increase the LM of smaller telescopes and expand their usage.

---

26 www.astrometrica.at

## 5.3 Comparing theoretical and observational results

In our measurements we were able to measure NEOs up to a magnitude of (19.50 ± 0.20) mag for GHOST and (18.00 ± 0.20) mag for ORT under ideal conditions with a stacked 4 × 60 s measurement. That is slightly above the range of the theoretical calculated values of 19.41 mag and 17.65 mag, respectively. However, the theoretical values refer to a single 60 s exposure, which means that the measured values are brighter than the expectation. Reasons for that might be found in the inaccuracies of some parameters used to estimate the performance (Sky Magnitude $m_{sky}$, noises $N_{bias}$, $N_{dark}$ and $N_{readout}$). Also, the total transmission $\tau$ of the telescope and the quantum efficiency $QE$ are based on the manufacturer's data for the camera, mirrors and filters might have decreased over the time. Also, to some extent, Eq. 11 is erroneous due to the assumption of an average wavelength $\bar{\lambda}$ of the incoming photons. Furthermore, we assumed in Eq. 12 an air mass $\chi = 1$, which is true for an object in the zenith, in fact objects have a lower altitude. Together, this can lead to a lower LM of the telescope.

According to the analysis of the NEOCC Priority List in Figure 8 with our estimation for the LM of our telescope in Figure 6 there are on average 9.2 observable NEOs per night in the Priority List for the GHOST telescope assuming a LM of 19.4 mag. In comparison we were able to submit on average 14.2 measurements of NEOs per night (Supplementary Table S1). With an average of three measurements per object per night, this results in 4.7 observed NEOs per night. Indeed, among the 27 NEO observation nights there were at least one third of the nights where the weather conditions were not clear for most of the night. Due to further limiting visual influences, such as the Moon, the faintest magnitude of the Sky Brightness could not be reached on many nights. Considering this, the theoretical expectation agrees with the results.

For other telescopes it needs to be considered that the obtained results are based on the parameters of the GHOST telescope, which are possibly only good assumptions for similarly structured telescopes. This means that the results only give estimates, but can they be calculated more precisely with the general Eqs 11, 12.

## 5.4 Outlook

A main feature that would be useful for our location is a real-time cloud detection. As we have stated, most nights have only short observation times due to clouds with rapidly changing weather conditions. A cloud-tracking software that is implemented in the Robotic Telescope system would maximize the possible observation time. Other optimizations can be made in the Scheduler making, such that objects are observed with the ideal observing conditions at the highest altitude possible and thus lower atmospheric extinction.

All in all, the system is useful especially considering an expected increase in the number of follow-up observations needed (Seaman et al., 2021). Currently, there are already not enough observations, and the need will increase by the accelerating rate of discovery caused by more survey telescopes.

# 6 Conclusion

We developed an optical telescope system with a fully robotic planning, scheduling and observation pipeline especially specialized for NEO observation, which is based on commercially available soft- and hardware components. This allows also other observatories to make use of these methods. Since its completion, it automatically generates measurements that are submitted to the MPC for improvements in trajectory predictions. A decisive improvement in the efficiency of observation time for already existing systems can be achieved by using the pipeline. However, with increasing need for follow-up observations it can also be easily used in the design of new follow-up telescopes.

# Data availability statement

The datasets presented in this study can be found in online repositories. The names of the repository/repositories and accession number(s) can be found in the article/Supplementary Material.

# Author contributions

TH contributed all plots and most text. TH and MG are the instrument scientists and the developer of the software. TP, GD, TO, and JK provided valuable feedback and reviewed the manuscript drafts. TP, JK, and BP are head of the facility. TS-R, MG, RR, and MZ provided external expertise and reviewed the manuscript draft. BP is the supervisor, contributed text, provided valuable feedback, and reviewed the manuscript drafts.

# Acknowledgments

We thank ESA's Near-Earth Object Coordination Centre for the provision of parts of the Priority List archive and especially we thank Detlef Koschny for the provided theoretical expertise. TS-R acknowledges funding from the NEO-MAPP project (H2020-EU-2-1-6/870377). This work was (partially) funded by the Spanish MICIN/AEI/10.13039/501100011033 and by "ERDF A way of making Europe" by the "European Union" through grant RTI2018-095076-B-C21, and the Institute of Cosmos Sciences University of Barcelona (ICCUB, Unidad de Excelencia "María de Maeztu") through grant CEX2019-000918-M.

# Conflict of interest

The authors declare that the research was conducted in the absence of any commercial or financial relationships that could be construed as a potential conflict of interest.

# Publisher's note

All claims expressed in this article are solely those of the authors and do not necessarily represent those of their affiliated organizations, or those of the publisher, the editors and the reviewers. Any product that may be evaluated in this article, or claim that may be made by its manufacturer, is not guaranteed or endorsed by the publisher.

# Supplementary material

The Supplementary Material for this article can be found online at: https://www.frontiersin.org/articles/10.3389/fspas.2022.895732/full#supplementary-material

# References

Boattini, A., D'Abramo, G., Valsecchi, G. B., and Carusi, A. (2007). A new protocol for the astrometric follow-up of near Earth asteroids. *Earth Moon Planets* 100, 31–41. doi:10.1007/s11038-006-9075-9

Bortle, J. E. (2001). Gauging light pollution: The bortle dark-sky scale. *Sky Telesc. Febr.*

Boslough, M., Brown, P., and Harris, A. (2015). "Updated population and risk assessment for airbursts from near-Earth objects (neos)," in 2015 IEEE Aerospace Conference, 1–12. doi:10.1109/AERO.2015.7119288

Bottke, W. F., Vokrouhlický, D., Rubincam, D. P., and Nesvorný, D. (2006). The yarkovsky and yorp effects: Implications for asteroid dynamics. *Annu. Rev. Earth Planet. Sci.* 34, 157–191. doi:10.1146/annurev.earth.34.031405.125154

Cibin, L., Chiarini, M., Gregori, P., Bernardi, F., Ragazzoni, R., Sessler, G., et al. (2019). "The fly-eye telescope, development and first factory tests results," in Proc. 1st NEO and Debris Detection Conference (ESA Space Safety Programme Office).

Crumey, A. (2014). Human contrast threshold and astronomical visibility. *Mon. Not. R. Astron. Soc.* 442, 2600–2619. doi:10.1093/mnras/stu992

Dotto, E., Banaszkiewicz, M., Banchi, S., Barucci, M. A., Bernardi, F., Birlan, M., et al. (2021). The EU project NEOROCKS — NEO Rapid observation, characterization, and key simulations project. *Eur. Planet. Sci. Congr.* 2021, EPSC2021–389. doi:10.5194/epsc2021-389

Falchi, F., Cinzano, P., Duriscoe, D., Kyba, C. C. M., Elvidge, C. D., Baugh, K., et al. (2016). The new world atlas of artificial night sky brightness. *Sci. Adv.* 2, e1600377. doi:10.1126/sciadv.1600377

García-Lozano, R., Rodes, J. J., Torrejón, J. M., Bernabéu, G., and Berná, J. Á. (2016). The busot observatory: Towards a robotic autonomous telescope. *Rev. Mex. Astron. Astrofísica Ser. Conf.* 48, 16–21.

Gupta, R., Singh, H. P., Kanbur, S. M., Schrimpf, A., and Dersch, C. (2015). U-smart - small aperture robotic telescopes for universities. *Publ. Korean Astronomical Soc.* 30, 683–685. doi:10.5303/PKAS.2015.30.2.683

Hänel, A., Posch, T., Ribas, S. J., Aubé, M., Duriscoe, D., Jechow, A., et al. (2018). Measuring night sky brightness: Methods and challenges. *J. Quantitative Spectrosc. Radiat. Transf.* 205, 278–290. doi:10.1016/j.jqsrt.2017.09.008

Hestroffer, D., Sánchez, P., Staron, L., Bagatin, A. C., Eggl, S., Losert, W., et al. (2019). Small solar system bodies as granular media. *Astron. Astrophys. Rev.* 27, 6. doi:10.1007/s00159-019-0117-5

Hodapp, K. W., Kaiser, N., Aussel, H., Burgett, W., Chambers, K. C., Chun, M., et al. (2004). Design of the pan-starrs telescopes. *Astron. Nachr.* 325, 636–642. doi:10.1002/asna.200410300

Holvorcem, P. R., Schwartz, M., Juels, C. W., Breganhola, M., Camargo, J., and Teixeira, R. (2003). Search by orcid, et alAstrometry of near-Earth asteroids using remotely-operated robotic telescopes. *Astronomy Lat. Am.* 1, 91–100.

Huebner, W. F., Johnson, L. N., Boice, D. C., Bradley, P., Chocron, S., Ghosh, A., et al. (2009). A comprehensive program for countermeasures against potentially hazardous objects (phos). *Sol. Syst. Res.* 43, 334–342. doi:10.1134/S003809460904008X

IAU General Assembly (2006). *Resolution b5: Definition of a planet in the solar system*. Busan, South Korea: IAU General Assembly.

Jurado Vargas, M., Merchán Benítez, P., Sánchez Bajo, F., and Astillero Vivas, A. (2002). Measurements of atmospheric extinction at a ground level observatory. *Astrophys. Space Sci.* 279, 261–269. doi:10.1023/A:1015184127925

Koschny, D., Drolshagen, E., Drolshagen, S., Kretschmer, J., Ott, T., Drolshagen, G., et al. (2017). Flux densities of meteoroids derived from optical double-station observations. *Planet. Space Sci.* 143, 230–237. doi:10.1016/j.pss.2016.12.007

Koschny, D., and Igenbergs, E. (2020). *Near-earth objects: A threat for earth? – or: Neos for engineers and physicists: Script for the course: Near-earth objects for engineers and physicists.*

Larson, S., Brownlee, J., Hergenrother, C., and Spahr, T. (1998). The catalina sky survey for neos. *Bull. Am. Astronomical Soc.* 30, 1037.

Lister, T. A., Greenstreet, S., Gomez, E., Christensen, E., and Larson, S. (2015). The lcogt neo follow-up network. *Proc. Int. Astron. Union* 10, 321–323. doi:10.1017/S1743921315006778

Mainzer, A., Abell, P., Bannister, M. T., Barbee, B., Barnes, J., Bell, J. F., III, et al. (2021). The future of planetary defense in the era of advanced surveys. *Bull. AAS* 53. doi:10.3847/25c2cfeb.ba7af878

Mamajek, E. E., Prsa, A., Torres, G., Harmanec, P., Asplund, M., Bennett, P. D., et al. (2015). "Iau 2015 resolution b3 on recommended nominal conversion constants for selected solar and planetary properties," in XXIX International Astronomical Union General Assembly (IAU).

Meftah, M., Damé, L., Bolsée, D., Hauchecorne, A., Pereira, N., Sluse, D., et al. (2018). Solar-iss: A new reference spectrum based on solar/solspec observations. *Astron. Astrophys.* 611, A1. doi:10.1051/0004-6361/201731316

Merline, W. J., and Howell, S. B. (1995). A realistic model for point-sources imaged on array detectors: The model and initial results. *Exp. Astron. (Dordr).* 6, 163–210. doi:10.1007/BF00421131

Micheli, M., Koschny, D., Drolshagen, G., Hainaut, O., and Bernardi, F. (2014). An esa neocc effort to eliminate high palermo scale virtual impactors. *Earth Moon Planets* 113, 1–13. doi:10.1007/s11038-014-9441-y

Micheli, M., Koschny, D., Drolshagen, G., Perozzi, E., and Borgia, B. (2015). Neo follow-up, recovery and precovery campaigns at the esa neo coordination centre. *Proc. Int. Astron. Union* 10, 274–281. doi:10.1017/S1743921315009175

Miguel, A. S. d., Zamorano, M. A. J., Kocifaj, M., Roby, J., and Tapia, C. (2017). Sky quality meter measurements in a colour changing world. *Mon. Not. R. Astron. Soc.* 467, 2966–2979. doi:10.1093/mnras/stx145

Milani, A. (1999). The asteroid identification problem. *Icarus* 137, 269–292. doi:10.1006/icar.1999.6045

Peña, J., Fuentes, C., Förster, F., Martínez-Palomera, J., Cabrera-Vives, G., Maureira, J. C., et al. (2020). Asteroids' size distribution and colors from hits. *Astron. J.* 159, 148. doi:10.3847/1538-3881/ab7338

Perna, D., Barucci, M. A., and Fulchignoni, M. (2013). The near-Earth objects and their potential threat to our planet. *Astron. Astrophys. Rev.* 21, 65. doi:10.1007/s00159-013-0065-4

Raab, H. (2002). "Detecting and measuring faint point sources with a ccd," in Meeting on Asteroids and Comets in Europe 2002.

Rumpf, C., Lewis, H. G., and Atkinson, P. M. (2016). The global impact distribution of near-Earth objects. *Icarus* 265, 209–217. doi:10.1016/j.icarus.2015.10.026

Seaman, R., Bauer, J., Brucker, M., Christensen, E., Grav, T., Jones, L., et al. (2021). NEO surveys and ground-based follow-up. *Bull. AAS* 53. doi:10.3847/25c2cfeb.9eb9da4e

Shporer, A., Brown, T., Lister, T., Street, R., Tsapras, Y., Bianco, F., et al. (2010). The LCOGT network. *Proc. Int. Astron. Union* 6, 553–555. doi:10.1017/s1743921311021193

Vereš, P., Payne, M. J., Holman, M. J., Farnocchia, D., Williams, G. V., Keys, S., et al. (2018). Unconfirmed near-Earth objects. *Astron. J.* 156, 5. doi:10.3847/1538-3881/aac37d